\documentclass[floatfix,twocolumn,rmp]{revtex4}

\usepackage{dcolumn,graphicx,amsmath,amssymb}
\usepackage{txfonts}

\begin{document}

\title{Korean university life in a network perspective:\\Dynamics of a
  large affiliation network}

\author{Petter Holme}
\affiliation{Department of Physics, University of Michigan, Ann Arbor,
  MI 48109}
\author{Sung Min Park}
\author{Beom Jun Kim}
\affiliation{Department of Molecular Science and Technology, Ajou
  University, Suwon 442-749, Korea}
\author{Christofer R. Edling}
\affiliation{Department of Sociology, Stockholm University, 10691
  Stockholm, Sweden}

\begin{abstract}
  We investigate course registration data of 18 semesters at a Korean
  University to portray the time evolution of students' positions in
  the network of fellow students. Apart from being a study of the
  social positions of students, the present work is also an example of
  how large-scale, time resolved, affiliation networks can be
  analyzed. For example we discuss the proper definitions of weights,
  and propose a redefined weighted clustering coefficient. Among other
  things, we find that the students enter the network at the center
  and are gradually diffusing towards the periphery. On the other
  hand, the ties to the classmates of the first semester (still
  present at the university) will, on average, become stronger as time
  progresses.
\end{abstract}

\maketitle

\section{Introduction}

Networks constitute, along differential equation models and cellular
automata, a fundamental framework for analyzing and modeling complex
systems~\cite{ba:rev,mejn:rev,doromen:book}. The advent of modern database 
technology has greatly vitalized the statistical study of
networks. Networks of electronic communication, genetic interaction,
hyperlinked web-pages, and so on, are available in sizes up to
hundreds of millions of vertices~\cite{broder:www}, to be compared
with the data sets of a decade, or so, ago that mostly were curated
manually edge by edge. The sizes of these computer compiled data
sets put new demands on algorithms and analysis methods---an $O(N^5)$
algorithm, where $N$ is the number of vertices, may work perfectly to
analyze a food-web~\cite{regeeco1} but would be intractable for
analysis of the contacts in an Internet community~\cite{pok}. On the
other hand, larger sizes rids the data of finite size biases, and
allows one to extrapolate the conclusions to the large size
limit. These new available data sets have created a new sub-field of
network-sociology~\cite{well:comp} and are the main reasons
statistical physicists (traditionally working in the large size limit)
have been joining this interdisciplinary field.

In the present paper we use course registration data from the mid-size
Korean University, Ajou University, located in Suwon, Republic of Korea.
Our data set consists of lists of undergraduate students registered to
courses for 18 semesters (two semesters per year), starting with the
spring semester 1996 and ending with the fall semester 2004. The basic
network of a particular semester is an ``affiliation
network'' where students and courses are two separate
sets of vertices and edges link students to courses to which they are
registered. From such a ``two-mode'' network (a network with two
classes of vertices) one can make a ``one-mode'' projection to the set
of students (or courses), where one student (course) is connected to
another if they have a link to the same course (student) in the
affiliation network. In this paper we only consider projections to the
set of students and try to answer how a student's time at the
university can be characterized by network statistics. Some basic
statistics of the network of the fall semester 2003 is given
in~\cite{our:ajounet}. Apart from allowing us to view Korean students'
time at the university, and maybe university students in general, we
introduce new structural quantities and methods of analyzing time
resolved affiliation network data.

\begin{figure*}
  \resizebox*{0.7\linewidth}{!}{\includegraphics{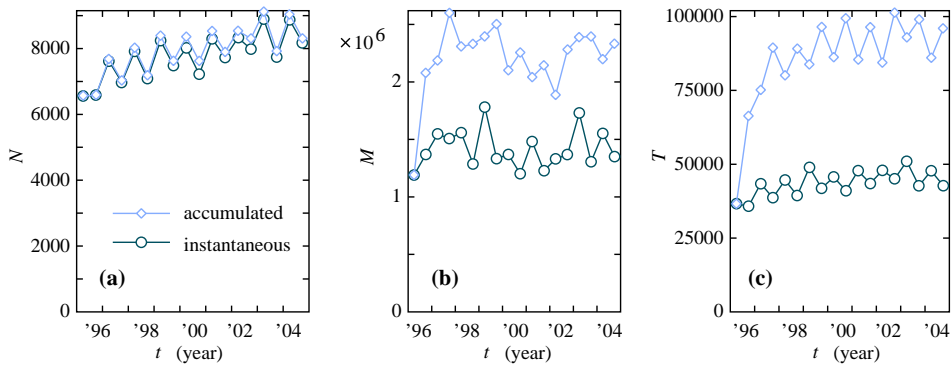}}
  \caption{The number of vertices $N$ (a), edges $M$ (b) and the total
    weight $T$ (c) in the
    networks. The instantaneous networks are the giant components of
  the projection of the course-student network for a given semester
  onto the set of students. The accumulated networks are giant
  components of the one-mode projection of accumulated course-student
  networks onto the sets of currently active students. An accumulated
  course-student network contains all courses in the data set given
  the specified semester or earlier.
  }
  \label{fig:size}
\end{figure*}

\begin{figure*}
  \resizebox*{0.7\linewidth}{!}{\includegraphics{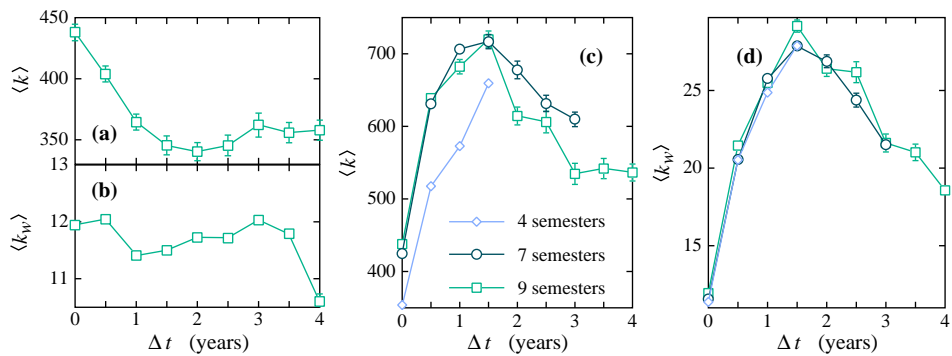}}
  \caption{
    The average degrees for students studying four, seven, and nine
  semesters. (a) and (b) show the unweighted and weighted results
  respectively for instantaneous networks. (c) and (d) show the
  unweighted and weighted results respectively for accumulated
  networks.
  }
  \label{fig:degwei}
\end{figure*}

\section{Construction of the networks}

From our lists of course-registration data we would like to
construct networks where an edge between two persons means that these
two students are likely to be acquainted. We will consider both
weighted and unweighted networks. For the weighted networks, we want a
high weight to represent a high probability of the two students being
acquainted. If two students take a course with few attendees they are
more likely to know each other than if the course is large. Following
Ref.~\cite{mejn:scicolpre2} we let a specific course $m$ contribute to
the weight between two students taking it with the inverse of the
total number of students taking $m$, $1/n(m)$. But if two students
take many courses in common, the chance of them being acquainted is
larger. To account for this we just sum the contribution of each
individual course
\begin{equation}\label{eq:w}
  w(i,j) = \sum_{m\in C} \frac{\delta_m(i,j)}{n(m)} ,
\end{equation}
where $C$ is the set of all courses, 
$\delta_m(i,j)$ is unity if both $i$ and $j$ take $m$ and $i\neq
j$, and zero otherwise. Now, as mentioned, our data is time resolved,
which gives us a possibility of accounting for the cessation of
acquaintances---two persons are less likely to feel acquainted if they
took a course together three years ago, than if they were in the same
class the last semester. The weight at a given time $t$ can thus be
written as
\begin{equation}\label{eq:w_t}
  w(i,j,t) = \sum_{t'\leq t}\sum_{m\in C}
  \frac{\rho(t-t')\:\delta_m(i,j,t')}{n(m,t')} ,
\end{equation}
where $\rho(t)$ is a non-increasing function that accounts for the decay
of friendships over time~\cite{my:ongoing}, and $n(m,t)$ is the number
of students taking the course $m$ at the semester $t$, and
$\delta_m(i,j,t)$ is the corresponding generalization of
$\delta_m(i,j)$. We will use the two simplest decay functions: Either
we let $\rho(t)$ decay as fast as possible and thus be unity for $t=0$ and
zero for $t>0$ giving an \textit{instantaneous} network, or we let
$\rho(t)$ not decay at all (be constantly unity) and obtain an
\textit{accumulated} network.

Some of the quantities we will use are based on shortest path
lengths. Since these are not defined in a general network we reduce
our graphs to their giant components (largest connected
subgraphs). In these, the numbers of vertices are 95-100\% of the
original networks. Another technicality concerns the temporal
boundaries---we do not know how long a student of the first semester
of our data has been at the university, and we do not know how long a
student present the last semester will stay. We handle this problem by
not including the students present the first and last semester in the
averages of our quantities. There are, of course, students who are on
leave from the university a few semesters. We estimate around $\sim
15\%$ of the data points are students who have returned from a
break. How to treat these students is a dilemma: On the one hand, one
would like the number of active semesters to be the measure of time (in
studying the time evolution of quantities). On the other hand, one
cannot exclude the returning students from the network, after all they
are a part of the network after their return. We will choose the latter
alternative, and treat a student on leave as still present at the
university, but with zero degree. The sizes of the networks are presented in
Fig.~\ref{fig:size}. We note that the accumulated networks sometimes
has a higher $N$ than the instantaneous counterparts. This is because
the additional edges of the accumulated networks can give a larger
giant component---the number of edges in the projection to the set of
students is the same. The jagged shape of Fig.~\ref{fig:size}(a) is an
effect of that more students enter the university the spring semester
($\sim 1957$ on average) than the fall semester ($\sim 353$ on
average). Assuming the half of students stay an odd (or even) number
of semesters the spring semesters would expect a difference of
$(1957-353)/2\approx 802$ between the spring and fall semesters. That
the actual average difference is smaller ($\sim 635$ on average), is
an effect of that more students take an even number of semesters since
many programs is for an even number of semesters.

\section{The measured statistics}

In this section we will present the quantities we measure, the values
we obtain and interpret these. All our curves are functions of the
time (number of semesters) $\Delta t$ a student has studied at the
university. To avoid, as much as possible, that behavioral
differences get averaged away, we measure the curves for students
present at the university a fixed time. As it turns out, these curves
are qualitatively similar (but sometimes differs quantitatively).

\subsection{Degree and weight\label{sec:degwei}}

A fundamental vertex-quantity of unweighted networks is the degree,
$k(i)$, defined as the number of edges leading to a vertex $i$. Degree
gives a measure of how central a vertex is in its local surrounding
(and is therefore sometimes referred to as \textit{degree
  centrality}~\cite{wf}). The straightforward generalization of degree
to weighted networks is~\cite{wf}:
\begin{equation}
  k_w(i)=\sum_j w(i,j)~.
\end{equation}
In Figs.~\ref{fig:degwei}(a) and (b) the degree, in its unweighted and
weighted versions, for instantaneous networks is plotted for students
staying nine semesters at the University. Other lengths of study give
the same general shape of both $\langle k\rangle$ and $\langle
k_w\rangle$---an early decrease of $\langle k\rangle$ that flattens
out, and a rather constant $\langle k_w\rangle$. This
decrease is probably a result of students taking increasingly
specialized courses, in smaller and smaller classes. Results for
accumulated networks are displayed in Figs.~\ref{fig:degwei}(c) and
(d). These have maximum around the fourth semester. Note that if none
of student A's fellow students of the first semester leave before A,
then A's degree would be strictly increasing in the accumulated
networks. The decreasing part is a result of old neighbors (vertices
one edge away) exiting the network.

\begin{figure}
  \resizebox*{0.85\linewidth}{!}{\includegraphics{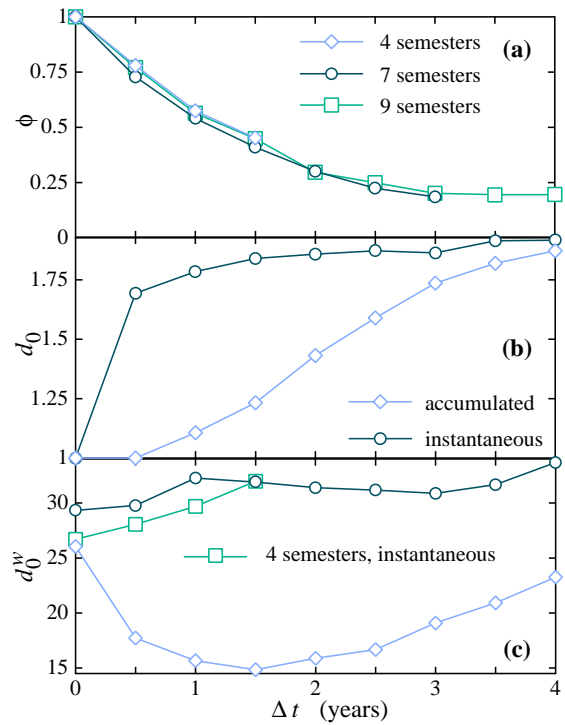}}
  \caption{
    Statistics about the fellow students of the first semester. (a)
    shows the fraction $\phi$ of students of the first semester that
    still are students of the university. (b) and (c) display the
    distance to the neighborhood of the first semester of students
    staying nine semesters in total (unless otherwise stated). (b)
    shows the unweighted version $d_0$ and (c) shows the weighted
    counterpart $d_0^w$. All errorbars are smaller than the symbol
    size.
  }
  \label{fig:dist}
\end{figure}

\subsection{Relation to fellow freshmen}

To get a picture of how the ties come and go in our networks, we study
the relation of a student to her, or his, fellow freshmen (defined as the
neighborhood of a student the first semester at the university). In
Fig.~\ref{fig:dist}(a) we display the average fraction of fellow
freshmen remaining after $t$ semesters $\phi$ for students who stay
four, seven, and nine semesters at the university. $\phi$ is
exceptionally similar for number of semesters at the university.

Students coming and leaving the University is of course not the only
dynamic mechanism present. The choice of courses can make students
drift apart or get closer. This can be measured by graph
distances. The  distance of a path $P$ between two students, in the
unweighted network, is the number of edges in it, whereas for the
weighted network it is the sum of reciprocal
weights~\cite{mejn:scicolpre2}:
\begin{subequations}
\begin{eqnarray}
d(i,j) & = & \min_{P\in\mathcal{P}(i,j)} |P|\\
d_w(i,j) & = & \min_{P\in\mathcal{P}(i,j)} \sum_{(i',j')\in P}
\frac{1}{w(i',j')}~,
\end{eqnarray}
\end{subequations}
where $\mathcal{P}(i,j)$ is the set of paths between $i$ and $j$.
In Fig.~\ref{fig:dist}(b) and (c) we plot the average shortest
distances between a nine-semester student and her, or his,
fellow freshmen still present in the network for unweighted $d_0$ and
weighted $d_0^w$ distances respectively. The unweighted distances are
strictly increasing. So one's fellow students of the first semester
does not only become fewer in number but also further away in
distances of binary networks. In the authors' personal experience, the
acquaintances with some of the fellow freshmen grew stronger with
time. This phenomenon is visible for the weighted distances of
Fig.~\ref{fig:dist}(c). Interestingly, $d_0^w$ of the instantaneous
network has both a maximum and a minimum in the interior of the
$\Delta t$ interval. We believe that this is because the students
choosing different courses than the nine-semester students (and thus
causing the early growth of $d_0^w$) are often leaving the university
sooner than the nine semester students. This picture is supported by
the monotonous growth seen for four-semester students. To calculate
weighted distances for all vertices is the computationally most
demanding part of our analysis, requiring $O(MN\log(N))$ time with a
Dijkstra's algorithm implemented using a binary heap for storing the
subsequently shortest distances~\cite{mejn:scicolpre2}.

\begin{figure}
  \resizebox*{0.85\linewidth}{!}{\includegraphics{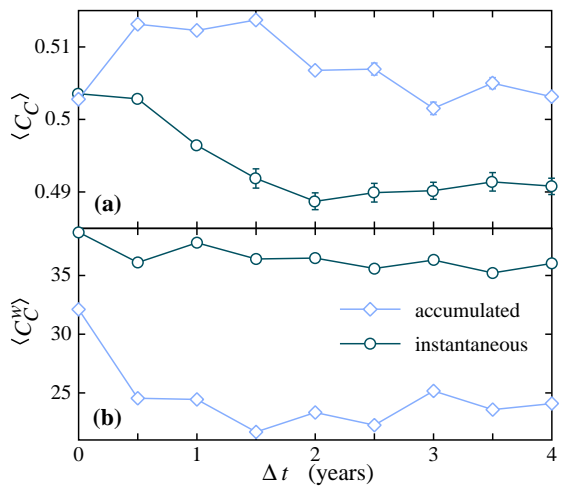}}
  \caption{
    Average closeness centralities of students who stay at the university
    nine semesters in total. (a) shows the unweighted version $d_0$ and
    (b) shows the weighted counterpart $d_0^w$.
  }
  \label{fig:clo}
\end{figure}

\subsection{Closeness centralities}

In Sect.~\ref{sec:degwei} we discussed the behavior of degree, a
quantity measuring the 
centrality of a student in the local surrounding. This section
concerns a global centrality measure. The study of global centrality
in graphs dates back to the 19th century~\cite{jordan:cent}. There are
many different notions of centrality, and thus, many different
measures, each trying to quantify a certain aspect of
centrality~\cite{wf}. One of the simplest is the \textit{closeness
  centrality}~\cite{beau:clo}, the reciprocal average shortest
distance to all other vertices:
\begin{equation}\label{eq:closeness}
  C_C(i) = \frac{N-1}{\sum_{j\neq i} d(i,j)} ~.
\end{equation}
(For weighted networks $d(i,j)$ is replaced by $d_w(i,j)$.) In
Fig.~\ref{fig:clo} we plot $C_C$ and $C_C^w$ for the nine-semester
students. The impression this figure gives us is a slowly decreasing
centrality. Other lengths of the university stays are not less shaky,
but all have a general downward trend. This leads us to the
conclusion, that as one starts the university, taking general courses
in big classes one is more central in the network of students than
when one takes more specialized courses later in one's university
education. It is worth noting that this holds even for the weighted
accumulated networks---it does not matter that one gets closer to
one's fellow freshmen (as discussed above), one still gets
increasingly peripheral in the network as a whole.
From the point of view of the educator, this analysis confirms what we
already know: that students are formative mainly in the first
semesters, when they are embedded in the short distances at the core
of the course-affiliation network. After the first semesters,
students' identities are already beginning to shape and you can merely
add to the already given foundation. From that point it is unusual
that students make dramatic shifts.

\begin{figure}
  \resizebox*{0.85\linewidth}{!}{\includegraphics{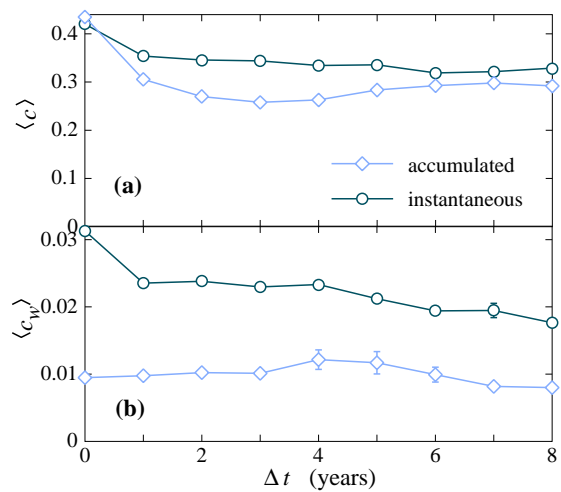}}
  \caption{
    Average clustering coefficient of students are a total of nine
    semesters at the University. (a) shows the unweighted version
    $\langle c\rangle$ and (b) shows the weighted counterpart $\langle
    c_w\rangle$. All errorbars are smaller than the symbol size.
  }
  \label{fig:clu}
\end{figure}

\subsection{Clustering coefficients}

Acquaintance networks are believed to have an overrepresentation of
strongly connected triads~\cite{holl:72,newco}. I.e., if A is strongly
connected to B and C, then B and C are also likely to have a strong
connection. A quantity to measure the strength of connections within $i$'s
neighborhood is the local clustering coefficient~\cite{wattsstrogatz}:
\begin{equation}\label{eq:cc}
  c(i) = e(\Gamma_i)\Big / \dbinom{k(i)}{2} ,
\end{equation}
where $e(\Gamma_i)$ is the number of edges within $i$'s
neighborhood $\Gamma_i$. The time evolution of clustering coefficients
for nine-semester students is plotted in Fig.~\ref{fig:clu}(a). We
believe the overall decrease reflects that the choice of courses a
student take is increasingly specialized and individual (which of
course decrease the fraction of fellow students themselves taking the
same course).

The decreasing sizes of the classes as a student progress to more
specialized subjects makes the weights between the students
stronger. Consequently one can argue that the triads become stronger
and the clustering coefficient should be adjusted to reflect this. It
turns out that the generalization if the local clustering coefficient
to weighted networks is not entirely trivial. We would like such a
measure to fulfill the following requirements:
\begin{enumerate}
\item The values should lie in the interval $[0,1]$.
\item If the weight matrix is replaced by a 0,1-adjacency matrix the
  Watts-Strogatz clustering coefficient should be recovered.
\item Zero weight should consistently represent the absence of an
  edge.
\item The contribution from one triad including the vertex $i$ to
  $i$'s weighted clustering coefficient $c_w(i)$ to be proportional to
  the weight of each edge in the triad.
\end{enumerate}
Unfortunately, none of the proposed definitions of a weighted
clustering coefficients we are aware
of~\cite{bar:wei,li:wei,onnela:wei} fulfills all these requirements. We
note that an alternative way of writing the Watts-Strogatz clustering
coefficient is \begin{equation}\label{eq:cc_alt}
c(i)=\frac{\sum_{jk}a_{ij}a_{jk}a_{ki}}
{\sum_{jk}a_{ij}a_{ki}} = \frac{\mathbf{A}^3_{ii}}
{(\mathbf{A}\mathbf{1}\mathbf{A})_{ii}}~,
\end{equation}
where $\mathbf{1}$ is the matrix where all elements are 1. In this
representation the generalization from an adjacency matrix to a weight
matrix is straightforward
\begin{equation}\label{eq:wcc}
c_w(i)=\frac{\sum_{jk}w_{ij}w_{jk}w_{ki}}
{\max_{ij}w_{ij}\sum_{jk}w_{ij}w_{ki}} = \frac{\mathbf{W}^3_{ii}}
{(\mathbf{W}\mathbf{W}_\mathrm{max}\mathbf{W})_{ii}}~,
\end{equation}
where $\mathbf{W}_\mathrm{max}$ is the matrix with $\max_{ij}w_{ij}$
on all positions, which indeed fulfills all four requirements
above. The weighted clustering coefficient counterparts of
Fig.~\ref{fig:clu}(a) is displayed in Fig.~\ref{fig:clu}(b). We see
that, for the instantaneous networks, the increasing strength of the
triads counterbalances the effect of more personalized curricula, so
the $\langle C_C^w\rangle$ curves are flatter.

\begin{figure}
  \resizebox*{0.85 \linewidth}{!}{\includegraphics{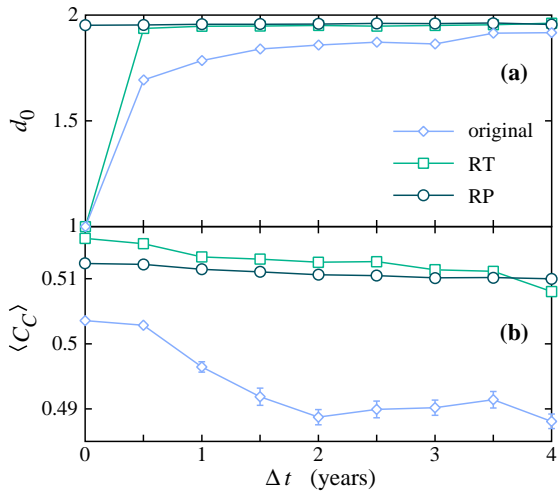}}
  \caption{Conditional uniform graph tests of $d_0$ (a) and $\langle
    C_C\rangle$ (b) for nine-semester students.
  }
  \label{fig:cug}
\end{figure}

\section{Conditional uniform graph tests}

In this section we put some the results of the measurements in
perspective by comparing them to results for graphs with some network
structure averaged away. A thorough analysis of this kind, to
establish the interpretations we propose above, would extend the size
of the paper so much that we omit it. Instead we perform case studies
of two of our unweighted quantities, $d_0$ and $\langle C_C\rangle$.

The standard way to view a complex network is to say it is to some
extent random and that it also has some degree of
structure---deviations from complete randomness induced by the forces
forming the network~\cite{fararo,rap:cont}. The commonly used
structural measures are not
independent. To sort out if a given quantity $X(G_0)$ (of a graph
$G_0$) is a dependent on a certain other quantity $Y(G_0)$ one can
perform a ``conditional uniform graph test''~\cite{katz:cug} and
compare $X(G_0)$ with the value of $X$ averaged over graphs with $Y$
conditioned to $Y(G_0)$. In this paper we will perform two simple
conditional uniform graph test: First we rewire the two-mode
course-student network before making the projection to the student
network, and constructing the weights (we call this RT randomization,
mnemonic for ``randomize two-mode''). We treat multiple course
registrations as a result of the randomization as just one course
registration. We also rewire the one-mode projections (RP, mnemonic
for ``randomize projection''). This gives Poisson random graphs of the
same sizes as the real networks. For this networks we average the
results over ten randomizations.

In Fig.~\ref{fig:cug}(a) we plot the average distance to the fellow
freshmen for the RT and RP randomizations. We see that the randomized
curves are approximately constant (apart from the RT curve where
$d_0(0)=1$ by definition). The distances in the real networks are
strictly shorter than in the randomized networks. This is completely
consistent with a picture of students of similar subject being closer
than students of different subjects---even if students are not
classmates after a few semesters they are likely to take courses of
similar subjects and thus be closer than an arbitrary other student.

Fig.~\ref{fig:cug}(b) shows the closeness centrality of nine-semester
students for real and randomized networks. We see that the students
have considerably lower centrality values than the vertices of the
test networks, and that the downward trend is stronger for the
original values. That the closeness is on average smaller in the real
networks reflects that the average distances are larger than in the
randomized networks. This is logical if one assumes the student
networks can be described as groups of students majoring in similar
subjects. Both randomizations remove the tendency of students close to
graduation take courses with few other students, which explains the
stronger decline of the real-world curves. The downward trend of the
randomized curves is explained by the increasing average distances of
the random networks due to the increasing number of students.

\section{Summary and discussion}

We have analyzed a data set from a Korean university consisting of
course registration lists for 18 semesters. These lists are made into
weighted and unweighted networks of students. The weight between two
students is a sum over all courses they have taken together. We argue
that, if one wants the weight to represent the probability of an
acquaintance along an edge, then the contribution to a weight from
one particular course should be chosen as inversely proportional to
the number of students taking the course. Furthermore, an old course
should contribute less to an edge than a newer. We use two decay
functions for the weights, one constant (minimally
decreasing)---defining accumulated networks, and one zero for any
course earlier than the present semester---defining instantaneous
networks. An unweighted edge is defined to be present whenever the
corresponding weight is non-zero. The quantities we measure, all in
both weighted and unweighted versions, and all as function of the time
a student has been present at the university are: Degree and closeness
centralities; distances to fellow students of the first semester; and
clustering coefficients. Some conclusions are strengthened by
conditional uniform graph tests.

We find that students enter the university in the center of the
student network, as a student progresses she, or he, will slowly
become more peripheral. On the other hand, the ties with fellow
students becomes stronger over time (to come to this conclusion one
needs weights, but the decay-function does not matter). The
connectedness of the neighborhood, as measured by the unweighted
clustering coefficient, decreases with time. We argue that this should
not be interpreted as the neighborhoods of students becomes weaker,
also here the conclusion from the weighted quantity is more
sound---that the triads of the instantaneous networks have roughly the
same strength over time.

From a qualitative point of view, interpreting our course-registration
network, and affiliation networks in general, is problematic in the
sense that it connects individuals with each other indirectly, and not
through directly observed social interaction. This means we do not
know if being in the same course also fosters significant social
interaction or if it breeds friendship for example. From personal
introspection only can we confirm that it sometimes did and it
sometimes did not. Still, it is perfectly reasonable to assume that
university course affiliation is an important identity shaper for a
university student, and therefore the analysis bears considerable
sociological relevance. In free societies most adult people join
schools, clubs, organizations, etc. by choice. However, once
affiliated one is under the influence of everything that goes on in
that particular arena~\cite{white:ic}. At the university, most, but not
all, of the students in a course are literary changing in front of our
eyes in the duration of a course. Indeed, that is one of the rewards
of teaching. And, once your students are finished with your course,
they go on to other courses that have been chosen by them partly
under the spell of whatever took place within the boundaries of that
particular course you were offering. In this respect the analysis
provides an illustration of the identity shaping processes that takes
place, not only in university courses but in every social arena with
which we are affiliated. Within the limits of course offerings, the
network is shaped by this identity seeking on behalf of the
students. Because individual identity is largely defined by exclusion
of other identities~\cite{simmel0,bourdieu} it is not surprising
that we see a growing fragmentation and distance between students as
time passes by. In network terms, university life as uncovered in the
analysis, begins at the core and drifts to the periphery. In terms of
identity shaping, the track is the opposite---students start off blank
in scholarly identity and gradually shapes into quasi-experts of their
majoring subject. And as the semesters pass, groups form around fellow
students that are following the same identity path, i.e.\ students
develop similar university identities.

\begin{acknowledgments}

We thank Ajou University for providing the course registration data
and Mark Newman for comments. CRE thank Wissenschaftskolleg zu
Berlin. Financial support from Korea Research Foundation through Grant
No.\ 2003-041-C00137 (BJK) and the Bank of Sweden Tercentenary
Foundation (CRE) are gratefully acknowledged.

\end{acknowledgments}

\end{document}